# Exploring the Urban - Rural Incarceration Divide: Drivers of Local Jail Incarceration Rates in the U.S.

Rachael Weiss Riley
Two Sigma Data Clinic
New York, NY, USA
rachael.riley@twosigma.com

Jacob Kang-Brown
Vera Institute of Justice
New York, NY, USA
jkangbrown@vera.com

Chris Mulligan
Two Sigma Data Clinic
New York, NY, USA
chris.mulligan@twosigma.com

Vinod Valsalam
Two Sigma Data Clinic
New York, NY, USA
vinod.valsalam@twosigma.com

Soumyo Chakraborty
Two Sigma Data Clinic
New York, NY, USA
soumyo.chakraborty@twosigma.com

Christian Henrichson
Vera Institute of Justice
New York, NY, USA
chenrichson@vera.org

**ABSTRACT**

As the rate of incarceration in the United States continues to grow, a large body of research has been primarily focused on understanding the determinants and drivers of federal and state prison growth. However, local jail systems, with 11 million admissions each year, have generated less research attention even though they have a far broader impact on communities. Preliminary time trend analysis conducted by the Vera Institute of Justice (Vera) uncovered disparities in county jail incarceration rates by geography. Contrary to assumptions that incarceration is an urban phenomenon, Vera discovered that during the past few decades, pretrial jail rates have declined in many urban areas whereas rates have grown or remained flat in rural counties. In an effort to uncover the factors contributing to continued jail growth in rural areas, Vera joined forces with Two Sigma's Data Clinic, a volunteer-based program that leverages employees' data science expertise. Using county jail data from 2000 – 2013 and county-specific demographic, political, socioeconomic, jail and prison population variables, a generalized estimating equations (GEE) model was specified to account for correlations within counties over time. The results revealed that county-level poverty, police expenditures, and spillover effects from other county and state authorities are all significant predictors of local jail rates. In addition, geographic investigation of model residuals revealed clusters of counties where observed rates were much higher (and much lower) than expected conditioned upon county variables.

## 1. INTRODUCTION
### 1.1. THE U.S. JAIL SYSTEM

The growth in the U.S. incarceration rate—which has more than quadrupled since the 1970s—is historically unprecedented and internationally unique.[1] Policymakers and the public, therefore, have been questioning whether prison and jail populations have grown too large. Reforms in the past decade have changed the trajectory of incarceration. The most recent data from 2015 shows that 1.5 million people are held under the jurisdiction of state and federal prisons, down from a peak of 1.6 million in 2009.[2] Local jail admissions numbered 11 million in 2015, down from 13.6 million in 2008.[3]

However, the more than 3,000 local jails have been studied far less than state prison systems, and many jails have continued to grow in recent years. State prisons hold people that have been convicted of crimes and sentenced to more than a year of punishment (in most cases). In contrast, 61% of the people in local jails on a given day (462,000 in 2013) have not been convicted.[1] Many individuals awaiting pretrial remain behind bars until their cases are resolved, merely because they cannot get the funds together for bail. Those sentenced to jail tend to be there for shorter stays – less than a year. Yet even a few nights in jail can have major impacts on an individual's family, housing arrangements and work.[4]

Given that the growth of prisons has been well documented, the Vera Institute of Justice launched the *Incarceration Trends Project* to analyze the growth of local jails, and found that the most dramatic growth and the highest jail incarceration rates were in unexpected places.

### 1.2. THE URBAN – RURAL INCARCERATION DIVIDE

Local jail populations grew from 157,000 on any given day in 1970, to over 700,000 people in 2015.[5] In the last 25 years, this growth was almost exclusively in pretrial detention, people that were not yet convicted of the charges they were facing.

The growth has a geographic dimension as well. Urban counties had the highest jail incarceration rates in 1970, but in more recent years, they tend to look more like their suburbs, and have lower jail incarceration rates. In contrast, while rural areas used to have relatively low rates of jail incarceration, in recent years their rates have been high and rising. As shown in Figure 1.2.1, the disparity between rural counties and metro areas has grown over time.

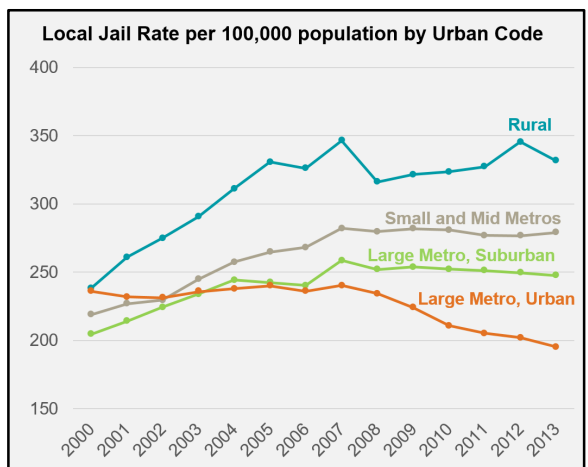

*Figure 1.2.1: Local jail rate per 100,000 population by urban code classification, 2000-2013*

While local jail rates were nearly identical for urban and rural counties in 2000, by 2013, rates in rural areas were 40% higher than those in urban metros.

| | # of counties | Millions of residents | % of population | People in jail | % of jail population |
|---|---|---|---|---|---|
| Large metro (Urban) | 62 | 95 | 31% | 203,143 | 27% |
| Large metro (Suburban) | 361 | 77 | 25% | 148,674 | 20% |
| Medium and small metro | 715 | 91 | 29% | 242,184 | 33% |
| Rural areas | 1,936 | 45 | 15% | 145,341 | 20% |
| Total | 3,074 | 308 | 100% | 739,342 | 100% |

Note: Excludes county data for six states (Alaska, Connecticut, Delaware, Hawaii, Rhode Island, and Vermont) that do not have local jails because there is a "unified" state prison-jail corrections system.

*Figure 1.2.2: Population and jail statistics for urban-rural counties, 2013*

In 2013, as Figure 1.2.2 demonstrates, rural counties had 15% of resident population, but 20% of nation's total jail population. This is surprising to many, and raises questions about what is driving the growth of jail incarceration, especially in smaller counties and rural areas.

## 1.3. VERA – TWO SIGMA DATA CLINIC PARTNERSHIP & RESEARCH GOALS

Vera's *Incarceration Trends Project* provided an opportunity to form a nonprofit – corporate collaboration to investigate an important social and political research topic using data science. Exploratory time trend analysis conducted by Vera researchers revealed a striking difference in local jail rates by county urbanicity. However, it was unclear what was driving this disparity. Why have jail rates declined in many urban areas while rising or remaining steady in smaller and more rural jurisdictions? To explore this question, Vera partnered with the Two Sigma Data Clinic, whose mission is to use data and technology to help nonprofits have a greater impact on the communities they serve. Over a series of meetings to align on project scope and vision, the following objectives were defined:

- To evaluate the characteristics of a county that are associated with local jail incarceration rates; and
- To identify counties with exceptionally high/low local jail rates conditioned upon observable characteristics.

## 2. METHODS
### 2.1. STUDY DATA
The data used in this study is part of the Incarceration Trends project. Vera developed the Incarceration Trends data tool so that Americans could have access to information showing how large their prisons and jails have grown, and who is held inside. Vera researchers compiled county-level historical data on jail populations from a variety of different public data sources, including the U.S. Department of Justice, Bureau of Justice Statistics (BJS) Census of Jails, covering all jails and conducted every five to eight years, and the Annual Survey of Jails, which covers about one-third of jails and has been conducted in non-census years since 1985.[6] This study focuses on jail data from 2000 onward to ensure measurement consistency across variables and to reduce missing data.

### 2.2 MEASURES
The outcome variable was local jail rates per 100,000 county population from 2000 to 2013 for U.S. Counties. Local jail population counts exclude individuals held in local jails on behalf of federal authorities like the U.S. Marshals or Immigration and Customs Enforcement (ICE), and represent the confined population on a given day, usually at the end of June.

Additional county-level measures describing the jail population included racial/ethnic composition (proportion non-Hispanic black jail inmates, proportion Hispanic jail inmates), percent awaiting trial, percent of inmates held for state authorities (usually for an overcrowded prison), and percent of inmates held for other counties (usually for an overcrowded jail). State-level measures included the state prison population per 100,000 and the proportion of a state's total prisoners held at local county jails.

Each county was assigned an "urban code" using a modified version of the 2013 National Center for Health Statistics Urban-Rural Classification Scheme for Counties. Counties were grouped into four categories using metropolitan statistical area delineation and population cutoffs- urban large metro (n=47), suburban large metro (n=338), small/mid-size metro (n=690), and rural (n=1,783). Large metros are those with more than 1 million residents. Rural areas are all counties outside of metropolitan areas.[7]

Additional county characteristics were obtained through a variety of public data sources and annual estimates for 2000 – 2013 were merged for each county-year when available. County demographics (total population, proportion non-Hispanic black residents and proportion Hispanic residents) were collected from the U.S. Census Bureau's Population and Housing Unit Estimates.[8] Socioeconomic data included poverty (population living below the federal poverty line) from the U.S. Census Small Area Income and Poverty Estimates and unemployment (percent of adults unemployed) from the U.S. Department of Labor Local Area Unemployment Statistics.[9,10] Data on county-level budget expenditures for welfare programs and police/corrections was acquired from the Government Finance Database.[11]

Due to missing data for some measures of a county-year observation, the final sample (n=33,616) included data for 2,858 counties, representing 93% of total counties with local jails.

## 2.3 STATISTICAL ANALYSIS

In order to account for the correlation of local jail rates within counties over time, a generalized estimating equation (GEE) was employed to evaluate the association between county, jail, and state characteristics and local jail rates. GEE extends generalized linear models to correlated non-normal outcomes:

$$\log(U(\beta)) = \sum_{i=1}^{N} \frac{\partial \mu_{ij}}{\partial \beta_k} V_i^{-1} \{Y_i - \mu_i(\beta)\}$$

Where the mean model $\mu_{ij}$ for county $i$ and year $j$ depends upon the regression parameters $\beta_k$ and variance structure $V_i$.

A series of nested log Poisson-GEE models were specified including bivariate associations, *urban code* only, *urban code + year*, and finally, *urban code + year + other characteristics*. An exchangeable working correlation structure was selected which assumes that within county correlations are consistent over time. Sensitivity analyses were conducted to test improvement in model fit under different working correlation structures. Model fit was evaluated using Pan's quasilikelihood information criterion (QIC) and the significance of each coefficient was tested using the Wald chi-square statistic.[12] Rate ratios (RR) represent exponentiated coefficients. Analyses were completed using the geepack library in R.[13,14]

## 3. RESULTS
### 3.1. NESTED GEE MODELS

Local jail rates averaged 284 per 100,000 population for years 2000 to 2013 and results from the bivariate analysis revealed that local jail rates vary significantly according to urban code classification. Local jail rates were, on average, 31.83% lower in suburban counties in large metropolitan areas (RR=0.68, 95%CI=0.62, 0.75) and 26.10% lower in urban counties in large metros (RR=0.74, 95%CI=0.67, 0.81) as compared to rural areas. The addition of year to control for global time trends strengthened the association between urban code and local jail rates. As shown in Figure 3.1.1, after accounting for year effects, local jail rates in rural counties were 34.62% higher than those in suburban metros (RR=0.65, 95%CI=0.59, 0.72), 31.38% higher than in urban metros (RR=0.69, 95%CI=0.62, 0.76), and 8.09% higher than those in small/mid-size metros (RR=0.92, 95%CI=0.85, 0.99).

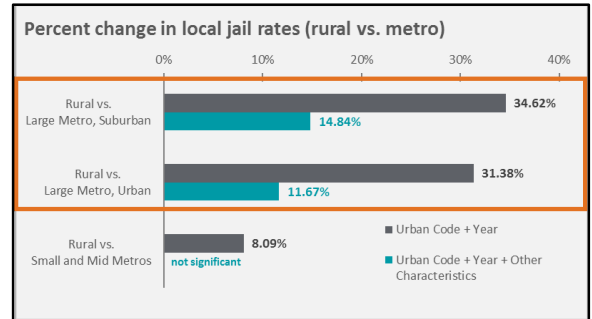

*Figure 3.1.1: Percent change in local jail rates for metro versus rural counties. Values represent the percent change in local jail rates from the referent category (rural) to each metro category (small/mid-size, urban, and suburban).*

After the inclusion of jail, county, and state covariates in addition to year effects, the magnitude of association between local jail rates and urban code was more than halved. Specifically, the percent decrease in local jail rates from rural counties declined from 34.62% to 14.84% in suburban metros and from 31.38% to 11.67% in urban metros, respectively. The difference in rates between small/mid- size metros and rural counties was no longer statistically significant.

Figure 3.1.2 depicts the percent change in local jail rates for every 10 percentage point increase in each jail, county, and state variable. County-level poverty had the greatest strength of association; local jail rates increased by 19.16% (CI=10.28%, 28.76%) for every 10 percentage point increase in poverty. Following poverty, significant relationships with local jail rates (in descending order of effect size) included the proportion of non-Hispanic black residents in the county (8.33%, CI= 4.12%, 12.70%), percent of jail inmates held under federal authority (7.76%, CI=5.65%, 10.01%), percent of jail population awaiting trial (-5.77%, CI= -6.97%, -4.55%), state prison rate (0.59%, CI= 0.37%, 0.81%), police and corrections expenditures (0.01%, CI= 0.0%, 0.01%), percent of jail inmates held under state authority (0.00%, CI= 0.00%, 0.01%), and percent of jail population held for other counties (0.00%, CI= 0.00%, 0.00%).

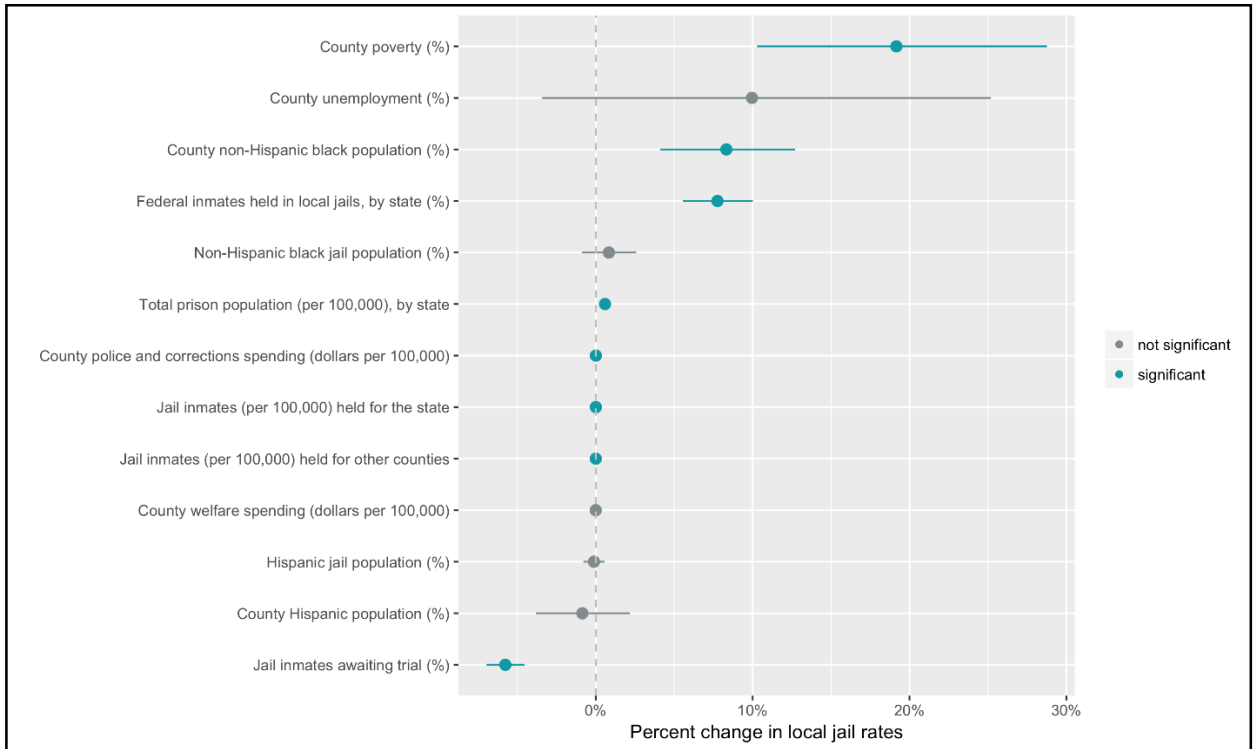

*Figure 3.1.2: Percent change in local jail rates for jail, state, and county measures. Values represent the percent change and 95% confidence interval in local jail rates for every 10 percentage point increase in each measure.*

County welfare expenditures, the proportion of non-Hispanic black and Hispanic jail inmates, the percent of unemployed county residents, and the proportion of Hispanic residents were all insignificant when they were included in the model. For the full results of the nested GEE models, see Table A.1.

## 3.2. ANALYSIS OF MODEL RESIDUALS

In an effort to identify specific counties and geographical clusters with exceptionally high (or low) local jail rates after conditioning upon included characteristics, model response residuals were calculated (observed − expected rates), mapped, and compared to raw local jail rates. Because measures that relate directly to public spending and criminal justice practice are especially interesting from a policy perspective, the residual model focuses on those variables and drops all race/ethnicity variables. Positive residuals signify that the observed local jail rate was higher than expected by the model; the converse is true for negative residuals.

Figure 3.2.1 demonstrates the geographic clustering of observed local jail rates by county. High rates are shown in darker pink and low rates in darker blue. It appears that local jail rates are not randomly distributed across space; evidence of geographic clustering is highlighted by yellow circles. Pockets of low rates appear in the northeast and midwest, whereas high rate clusters are visible in Florida and Utah. The presence of hot and cold spots indicate that drivers of local jail rates are likely operating at a more local or perhaps regional level.

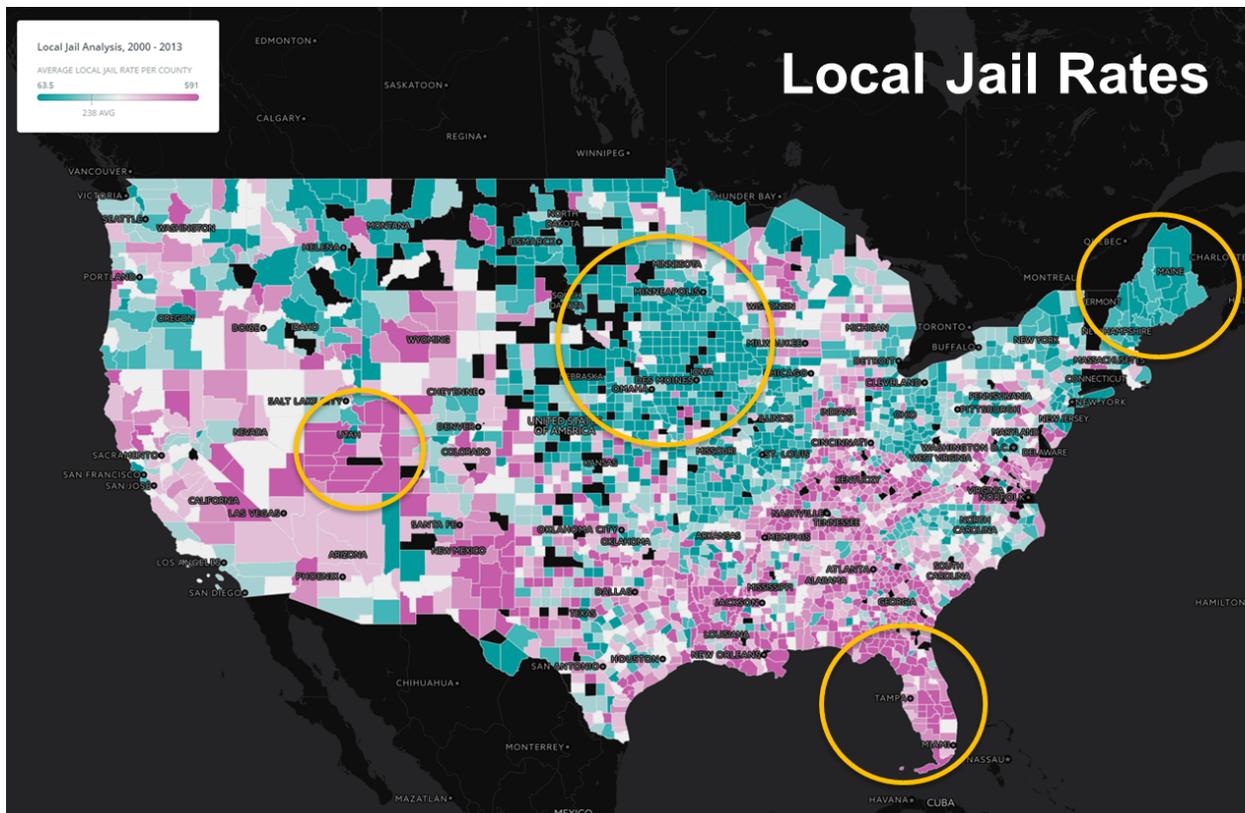

*Figure 3.2.1: Geographic clustering of average local jail rates by county, 2000 - 2013. Yellow circles represent high or low rate clusters*

Comparing the map of local jail rates to that of the model residuals (see Figure 3.2.2), it is apparent that the geographic clusters of high and low rates are not explained away by the jail, state, and county characteristics included in the model. Yellow circles depict clusters that persist after controlling for additional measures, and green circle represent the appearance of new groupings. With the exception of the midwest (red circle), it is clear that the explanatory variables included in the model do not reduce the spatial patterning of local jail rates by county, suggesting that unknown or unmeasurable factors are operating at the cluster-level. In addition, cluster boundaries appear to sometimes fall along state lines, indicating the presence of state level characteristics and policies that aren't included in the model.

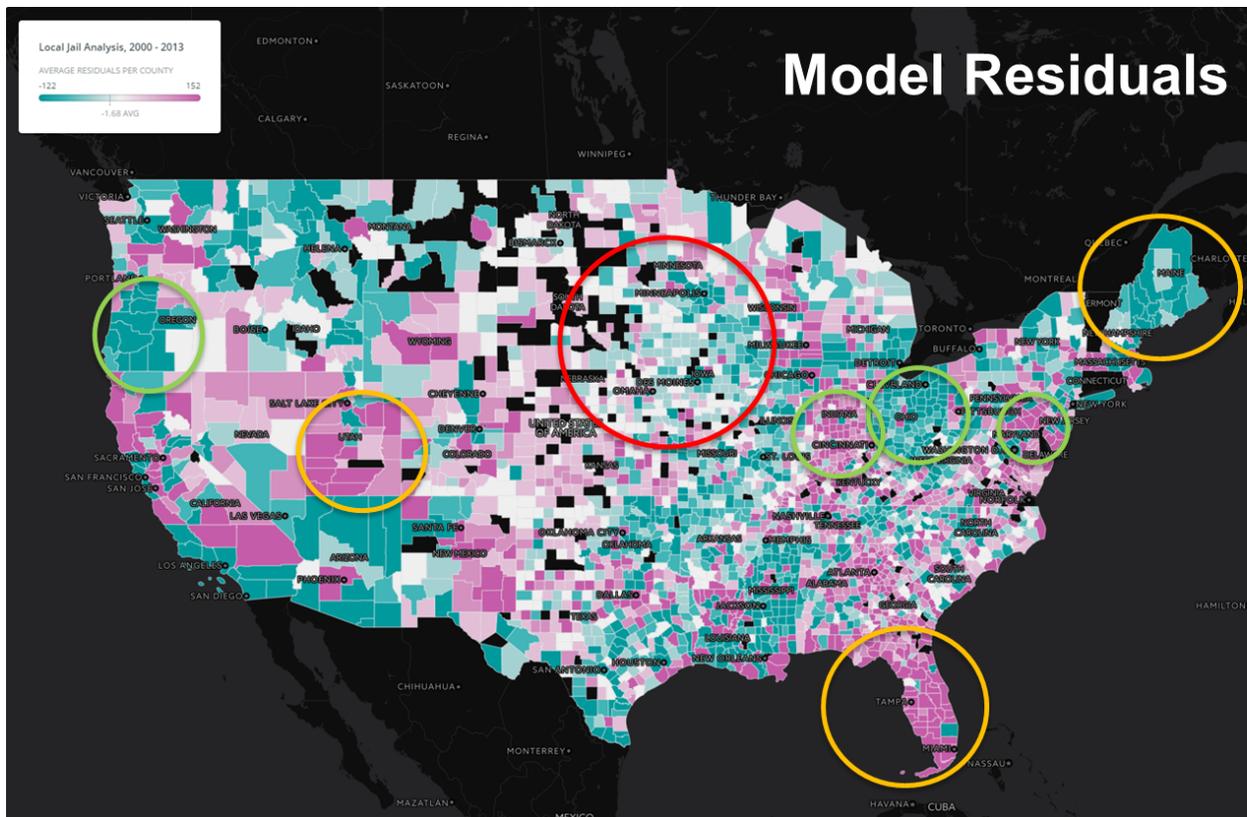

*Figure 3.2.2: Geographic clustering of average GEE model residuals, 2000 - 2013. Yellow circles represent persistent clusters of high and low rates, green depict new groupings, and red highlights the disappearance of a cluster.*

## 4. DISCUSSION AND NEXT STEPS

According to the results of this study, more than half of the variation in local jail rates by urban code was explained by the covariates included in the model. Findings revealed that poverty, demographics, police and corrections expenditures, and spillover effects from other county and state authorities were all significantly associated with local jail rates.

Much of the urban-rural disparity in jail rates was accounted for by county-level poverty. Poverty can influence the size of a jail in a number of ways. First, places with higher rates of poverty often struggle to provide government services, which includes the many functions that are necessary to process court cases—and when cases move slowly, the jail population grows. Second, poorer counties often cannot afford jail diversion and drug treatment programs, which many large cities have adopted, because the upfront investment is prohibitively expensive, even if it yields long-term jail savings. Third, considering that a large proportion of people in local jails are awaiting trial, poverty is likely to directly influence the ability of these individuals to pay bail, thus influencing local jail rates.[5]

The relationship between poverty and local jail is further supported by the results of a truncated model (not reported in this paper) that included county-level burglary/robbery crime and arrest rates. In this model, poverty had the greatest association with local jail rates, whereas crime and arrests were insignificant. Because crime reporting practices change over time and across jurisdictions, the most appropriate measures are those that closely align with estimates from the National Crime Victimization Survey, namely burglary and robbery.[15] This suggests a possible disconnect between what might be assumed to drive local jail rates (i.e., crime and arrests) versus what is actually influencing them (i.e., poverty) and warrants additional research.

Geographic investigation of model residuals revealed clusters of counties where observed rates were much higher (or much lower) than expected conditioned upon county variables included in the model. The spatial patterning evident in both local jail rates and model residuals indicates the presence of unmeasured factors that are operating at the local, state, or regional level. For instance, higher jail rates in California could be attributed to the expanded role for jails to hold sentenced individuals after state policy changed in 2009.

### 4.1. LIMITATIONS

Jail survey data have limitations and potential bias. Data collection is carried out at the county-level, and data errors and incorrect entries are hard to control. Although this survey is conducted bi-annually, and the presence of multiple years of data helps to reduce these errors, the presence of erroneous data cannot be ruled out. In addition, some county-year observations were not included in the analysis due to missing data, which could have introduced some bias. Lastly, the outcome measure was a composite that included the pretrial and local county population as well as the jail population held under state and other county authority, which may be less consistently measured across states.

## 4.2. NEXT STEPS AND POLICY CONCLUSIONS

In order to address some of these limitations, future research should use alternative data that might be more comparable across states, such as pretrial jail population numbers. Pretrial detention is a core function of jails and reflects the operation of the local criminal justice system— whereas the variation between sentenced populations in jails appears to reflect state level policies—in some states, longer sentences may be served in local jails. The jail population used in the rates for this analysis is a combination of pretrial and sentenced populations, which may impact the analysis. Above and beyond this specific issue, there may be other unobserved factors, such as sentencing and bail laws, that vary systematically by state and should be explored in future research. This additional research could be used to establish spatial clusters with high or low rates that persist across model specifications. In these clusters, other research methods—qualitative research, investigative journalism, or historical analysis could illuminate this further.

## 5. ACKNOWLEDGEMENTS

We thank the entire Vera team for the excellent work they do in the pursuit of equal justice, ending mass incarceration, and strengthening communities. We would like to acknowledge the broader Two Sigma Data Clinic team—Christine Zhang, Roxanne Zalucky, Jeffrey Saret, Ben Wellington, Greg Shih, Jim Charatan, Ris Sawyer, Katy Knight, Dave Hahn, and Thea Charles—for their expertise, support, and feedback on the project and manuscript. Lastly, a special thanks to Katy Knight for introducing this partnership.

## 6. DISCLAIMER

# APPENDIX

Table A.1: Nested GEE model results for local jail rate per 100,000 population, 2000 - 2013

|  |  | Urban Code | | | | Urban Code + Year | | | | Urban Code + Year + Other Characteristics | | | |
|---|---|---|---|---|---|---|---|---|---|---|---|---|---|
|  |  | Estimate | Std.err | Rate Ratio (95% CI) | Pr(>\|W\|) | Estimate | Std.err | Rate Ratio (95% CI) | Pr(>\|W\|) | Estimate | Std.err | Rate Ratio (95% CI) | Pr(>\|W\|) |
|  | (Intercept) | 5.59 | 0.03 |  | 0.0000 | 5.53 | 0.03 |  | 0.0000 | 4.95 | 0.08 |  | 0.0000 |
| Urban Code | URBAN_CODE: Rural* |  |  |  |  |  |  |  |  |  |  |  |  |
|  | URBAN_CODE: Large Metro, Suburban | -0.38 | 0.05 | 0.68 (0.62, 0.75) | 0.0000 | -0.42 | 0.05 | 0.65 (0.59, 0.72) | 0.0000 | -0.16 | 0.05 | 0.85 (0.77, 0.94) | 0.0012 |
|  | URBAN_CODE: Large Metro, Urban | -0.30 | 0.05 | 0.74 (0.67, 0.81) | 0.0000 | -0.38 | 0.05 | 0.69 (0.62, 0.76) | 0.0000 | -0.12 | 0.06 | 0.88 (0.79, 0.99) | 0.0375 |
|  | URBAN_CODE: Small and Mid Metros | -0.06 | 0.03 | 0.95 (0.88, 1.01) | 0.1105 | -0.08 | 0.04 | 0.92 (0.85, 0.99) | 0.0240 | 0.04 | 0.03 | 1.04 (0.98, 1.10) | 0.2191 |
| Year | YEAR: 2000* |  |  |  |  |  |  |  |  |  |  |  |  |
|  | YEAR: 2001 |  |  |  |  | 0.01 | 0.01 | 1.01 (1.00, 1.03) | 0.0707 | 0.01 | 0.01 | 1.01 (0.99, 1.03) | 0.2799 |
|  | YEAR: 2002 |  |  |  |  | 0.03 | 0.01 | 1.03 (1.02, 1.05) | 0.0001 | 0.01 | 0.01 | 1.01 (0.98, 1.04) | 0.3919 |
|  | YEAR: 2003 |  |  |  |  | 0.07 | 0.01 | 1.07 (1.05, 1.09) | 0.0000 | 0.05 | 0.02 | 1.05 (1.01, 1.09) | 0.0142 |
|  | YEAR: 2004 |  |  |  |  | 0.10 | 0.01 | 1.10 (1.08, 1.13) | 0.0000 | 0.07 | 0.02 | 1.07 (1.04, 1.10) | 0.0000 |
|  | YEAR: 2005 |  |  |  |  | 0.13 | 0.01 | 1.14 (1.11, 1.17) | 0.0000 | 0.09 | 0.01 | 1.10 (1.07, 1.13) | 0.0000 |
|  | YEAR: 2006 |  |  |  |  | 0.14 | 0.02 | 1.15 (1.12, 1.18) | 0.0000 | 0.09 | 0.02 | 1.09 (1.06, 1.13) | 0.0000 |
|  | YEAR: 2007 |  |  |  |  | 0.15 | 0.01 | 1.16 (1.13, 1.19) | 0.0000 | 0.09 | 0.02 | 1.09 (1.06, 1.13) | 0.0000 |
|  | YEAR: 2008 |  |  |  |  | 0.14 | 0.01 | 1.15 (1.12, 1.18) | 0.0000 | 0.06 | 0.02 | 1.06 (1.02, 1.11) | 0.0020 |
|  | YEAR: 2009 |  |  |  |  | 0.12 | 0.02 | 1.12 (1.09, 1.16) | 0.0000 | -0.01 | 0.04 | 0.99 (0.92, 1.07) | 0.8092 |
|  | YEAR: 2010 |  |  |  |  | 0.09 | 0.02 | 1.09 (1.06, 1.13) | 0.0000 | -0.04 | 0.04 | 0.96 (0.89, 1.04) | 0.2902 |
|  | YEAR: 2011 |  |  |  |  | 0.06 | 0.02 | 1.07 (1.03, 1.11) | 0.0007 | -0.09 | 0.04 | 0.92 (0.85, 0.99) | 0.0247 |
|  | YEAR: 2012 |  |  |  |  | 0.07 | 0.02 | 1.07 (1.03, 1.11) | 0.0009 | -0.07 | 0.03 | 0.93 (0.87, 1.00) | 0.0436 |
|  | YEAR: 2013 |  |  |  |  | 0.06 | 0.02 | 1.06 (1.03, 1.10) | 0.0010 | -0.05 | 0.03 | 0.95 (0.89, 1.02) | 0.1324 |
| Jail | JAIL_LATINO_PERCENT |  |  |  |  |  |  |  |  | 0.00 | 0.00 | 1.00 (1.00, 1.00) | 0.7388 |
|  | JAIL_BLACK_PERCENT |  |  |  |  |  |  |  |  | 0.00 | 0.00 | 1.00 (1.00, 1.00) | 0.3402 |
|  | JAIL_PRETRIAL_PERCENT |  |  |  |  |  |  |  |  | -0.01 | 0.00 | 0.99 (0.99, 1.00) | 0.0000 |
|  | JAIL_OTHERCOUNTIES_RATE |  |  |  |  |  |  |  |  | 0.00 | 0.00 | 1.00 (1.00, 1.00) | 0.0000 |
|  | JAIL_STATES_RATE |  |  |  |  |  |  |  |  | 0.00 | 0.00 | 1.00 (1.00, 1.00) | 0.0000 |
| County | HISPANIC_PERCENT |  |  |  |  |  |  |  |  | 0.00 | 0.00 | 1.00 (1.00, 1.00) | 0.5762 |
|  | NHBLACK_PERCENT |  |  |  |  |  |  |  |  | 0.01 | 0.00 | 1.01 (1.00, 1.01) | 0.0001 |
|  | POVERTY_PERCENT |  |  |  |  |  |  |  |  | 0.02 | 0.00 | 1.02 (1.01, 1.03) | 0.0000 |
|  | UNEMPLOYMENT_PERCENT |  |  |  |  |  |  |  |  | 0.01 | 0.01 | 1.01 (1.00, 1.02) | 0.1515 |
|  | WELF_EXP_RATE |  |  |  |  |  |  |  |  | 0.00 | 0.00 | 1.00 (1.00, 1.00) | 0.1476 |
|  | POLICE_EXP_RATE |  |  |  |  |  |  |  |  | 0.00 | 0.00 | 1.00 (1.00, 1.00) | 0.0000 |
| State | PRISON_JAIL_PERCENT |  |  |  |  |  |  |  |  | 0.01 | 0.00 | 1.01 (1.01, 1.01) | 0.0000 |
|  | PRISON_TOTAL_RATE |  |  |  |  |  |  |  |  | 0.00 | 0.00 | 1.00 (1.00, 1.00) | 0.0000 |
|  | N |  |  | 33616 |  |  |  | 33616 |  |  |  | 33616 |  |
|  | Clusters |  |  | 2858 |  |  |  | 2858 |  |  |  | 2858 |  |
|  | Model fit: QIC |  |  | -89246363 |  |  |  | -89240954 |  |  |  | -90183548 |  |

*Referent; QIC = quasi likelihood information criteria*